 \pdfoutput=1 

\documentclass[conference]{IEEEtran}
\ifCLASSINFOpdf
\else
\fi

\usepackage{times}
 \usepackage{helvet}
  \usepackage{courier}
  \usepackage[hyphens]{url}
  \usepackage{graphicx}
  \usepackage{adjustbox}
  \usepackage{graphicx}
\usepackage{amsmath,amssymb,amsfonts}
\usepackage{algorithmic}
 \usepackage{comment}
 \usepackage{todonotes}
 \usepackage{color}
 \usepackage{url}
\usepackage{amsmath}
\ifCLASSOPTIONcompsoc
    \usepackage[caption=false, font=normalsize, labelfont=sf, textfont=sf]{subfig}
\else
\usepackage[caption=false, font=footnotesize]{subfig}
\fi
\newtheorem{definition}{Definition}
\newtheorem{theorem}{Theorem}
\newtheorem{corollary}{Corollary}[theorem]
\newtheorem{lemma}[theorem]{Lemma}

\newcommand{\qed}{\hfill\blacksquare}

\begin{document}
\pdfinfo{
  /Title (Fairness Perception from a  Network-Centric Perspective)
  /Author (Submitted for blind review)
  /Keywords (Fairness, Link prediction, network mining)  
  }
\title{Fairness Perception from a  Network-Centric Perspective}
\title{Fairness Perception from a  Network-Centric Perspective}
\author{\IEEEauthorblockN{Farzan Masrour, Pang-Ning Tan, Abdol-Hossein Esfahanian}
\IEEEauthorblockA{\textit{Department of Computer Science and Enginering, Michigan State University}\\
Emails:\{masrours, ptan, esfahanian\}@cse.msu.edu}
}

\maketitle

\begin{abstract}
Algorithmic fairness is a major concern in recent years as the influence of machine learning algorithms becomes more widespread. In this paper, we investigate the issue of algorithmic fairness from a network-centric perspective. Specifically, we introduce a novel yet intuitive function known as network-centric fairness perception and provide an axiomatic approach to analyze its properties. Using a peer-review network as case study, we also examine its utility in terms of assessing the perception of fairness in paper acceptance decisions. We show how the function can be extended to a group fairness metric known as fairness visibility and demonstrate its relationship to demographic parity. We also illustrate a potential pitfall of the fairness visibility measure that can be exploited to mislead individuals into perceiving that the algorithmic decisions are fair. We demonstrate how the problem can be alleviated by increasing the local neighborhood size of the fairness perception function.  
\end{abstract}


%
\IEEEpeerreviewmaketitle

\section{Introduction}
The influence of machine learning algorithms is pervasive across numerous applications, from healthcare and e-commerce to financial and criminal justice systems. Despite their utility, previous studies have shown that 
these algorithms may inherently contain unintended biases that discriminate against certain groups of the population. For instance, an investigation into the COMPAS software \cite{angwin2016machine}, which is used to  predict recidivism among past offenders, was found to be biased against African Americans. Concerns have also been raised on the unfair treatment of various marginalized groups in other applications such as online advertising \cite{datta2015automated}, mortgage application \cite{corbett2017algorithmic}, and predictive policing \cite{jefferson2018predictable,brantingham2018does}. 

The challenge of removing biases and discrimination from algorithm decision-making process has gained tremendous attention in recent years. A significant amount of efforts have been devoted towards defining the notion of algorithmic fairness.  To date, various mathematical formulations of fairness have been proposed. For instance, group fairness definitions  such as demographic parity and equalized odds have been developed to assess the degree of prejudice against certain protected groups in the population. 
While each of these definitions has its own merits, it has been shown that many of them are incompatible with each other  \cite{chouldechova2017fair,friedler2016possibility}. 

The group fairness measures are designed to determine the level of equity among different groups of individuals who are either harmed by or benefited from the algorithmic-driven decisions.  
However, the consequences of 
an unfair decision may extend beyond those individuals who are directly impacted by the decision. In fact, they may elicit negative responses from other individuals who identified themselves to be in the same group as the affected individuals.  
For instance, hiring discrimination 
against a qualified member from an underrepresented group not only affects the well being of that individual, but will also have an adverse effect on other members of the underrepresented group who observed such a behavior. This example suggests that fairness assessment must take into consideration the perception of other individuals who may not be directly impacted by the algorithmic decisions. This has led to the growing interest to understand perceived fairness in individuals \cite{peiro2014,shulga2018}.

\begin{figure}[t]
    \centering
    \includegraphics[scale=0.25]{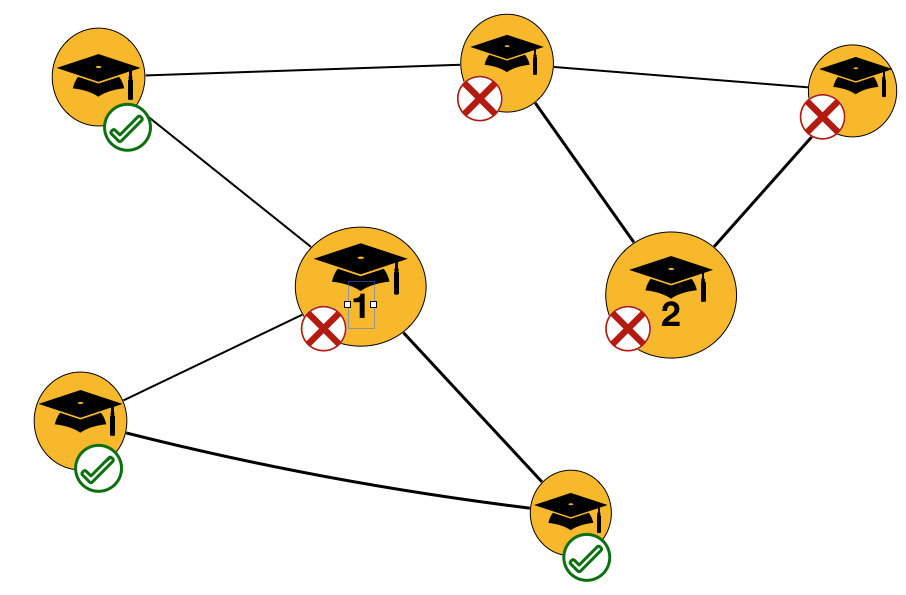}
    \caption{An example illustration of network-centric fairness perception.}
    \label{fig:example}
\end{figure}

Fairness perception is rooted in social comparison theory. For instance, equity theory~\cite{adams1965inequity} argues  that ``humans do not base their satisfaction on what they receive but rather what they receive in relation to what they think they should receive". The reaction of an individual to the outcome of a decision process is based on the expectation of the individual and this expectation not only depends on one's own outcome  but also the outcomes of other individuals they are aligned with, which we refer to as the \emph{reference group}.  The choice of the reference group is typically influenced by the similarity measure we use, e.g., we may compare ourselves to our co-workers, friends, and family members. By observing the outcomes of other members in our reference group, this will help shape our expectation about what should be considered a fair outcome. For instance, consider the salary hike decisions for employees in an organization. The perception of fairness is often related to comparing the current salary of an employee to the pay scale of other employees in the same reference group (i.e., those who are doing similar jobs in the same or in other organizations). 

In this paper, we examine the notion of fairness perception from a network analysis perspective. Networks provide a natural way to represent individuals and their connections to other individuals in the same reference group.
For instance, Figure \ref{fig:example} shows a toy example of a network of students applying for college admission to a prestigious university. In this network, two students are linked if they know each other. Suppose the admission committee of the college has decided to accept 3 of the applicants, denoted  as nodes with green check marks, and to reject the other 4 applicants. For brevity, we assume all the students have similar qualifications. Consider the two students labeled as 1 and 2, respectively. 
Although both applicants were rejected, their expectations for admission and perceptions of fairness are very different. 
Student 1 has a higher expectation of being admitted compared to student 2 since all of his/her friends were accepted. Thus, the perception of fairness for student 1 is different than that for student 2.

In this paper, we introduce the notion of network-centric fairness perception and illustrates its potential application in the context of peer review process. Peer evaluation of scientific work has significant effect on scientific advancement. However, similar to other systems designed by humans, it is potentially biased, favoring certain groups of individuals (e.g., famous authors or researchers from well-known institutions) \cite{stelmakh2018peerreview4all,tomkins2017reviewer}. The unintended consequences of the bias in peer review go beyond the decision about a single paper, it can significantly effect the career path of the authors. Indeed, with the explosion in the number of submitted papers for conferences and journals, the need to improve the peer review process is more important than ever. In this study, we illustrate how the notion of fairness perception can be applied to a peer review network of scientific papers submitted to a major computer science conference, where the links between two papers are established based on their co-author relationships. We show how the proposed network-centric fairness perception function can be used to assess the perception of authors about the paper acceptance decisions. An axiomatic approach for analyzing properties of fairness perception functions is also presented. We then extend our network-centric perception function to a group fairness measure known as fairness visibility and show its relationship to demographic parity under certain mild assumptions. We also describe a potential pitfall of assessing fairness from a local neighborhood perspective. Specifically, it can  mislead individuals into thinking that the decision-making process is fair even though the decisions are still biased toward certain groups of nodes. We show how the problem can be alleviated by expanding the local neighborhood size of the fairness perception function.


The main contributions of this paper are as follows:
\begin{itemize}
    \item We present a novel approach for measuring individual fairness based on a novel network-centric fairness perception function.
    \item We provide theoretical analysis on properties of the fairness perception function.
    \item We show how the individual fairness perception function can be extended to group fairness.
    \item We present a case study based on peer review network to demonstrate the efficacy of our proposed fairness perception function and fairness visibility measures.
\end{itemize}

\section{Related Work}

The subject of algorithmic fairness has received considerable attention among  researchers in recent years~\cite{berk2018fairness,dwork2012fairness,hardt2016equality,kusner2017counterfactual}. Previous works have mostly focused on algorithmic fairness for independent and identically distributed (i.i.d.) data. These previous works can be classified into two categories---individual and group fairness. The notion of fairness at individual level is based on the premise that similar people should be treated similarly~\cite{dwork2012fairness}, with similarity being defined based on a task-specific metric. 
As another example, Joseph et al.~\cite{joseph2016fairness} examined individual fairness in multi-armed bandit problems, by requiring that unqualified individuals should never be rewarded at the expense of more qualified ones. 

For group fairness, the individuals are partitioned into groups (e.g., based on some protected attributes) and the parity
of a statistical measure defined on the groups is used to assess fairness. These measures are typically computed from a confusion matrix~\cite{berk2018fairness} with the goal of ensuring that the average value of the measure does not vary significantly across different groups~\cite{feldman2015certifying,hardt2016equality,dwork2012fairness,feldman2015certifying}.  
A widely used criterion belongs to this group is \textit{demographic parity} or \textit{statistical parity}~\cite{dwork2012fairness}, which 
constrains the output of classification problem to be independent of the protected attribute. 
Demographic parity has been used in a number of previous papers
\cite{louizos2015variational,kamishima2011fairness,johndrow2019algorithm,edwards2015censoring,calders2009building}, which considers the degree of independence between the model output and protected attribute. 
Another well known group fairness measure is \textit{equalized odds} ~\cite{hardt2016equality}, which seeks to ensure that the predictions are conditionally independent of the protected group classification task. This family of fairness definitions are simple and can be achieved without making any assumptions about the data. However, recent studies have raised concerns about the fairness guarantees achieved by the individual or group fairness measures. 
First, the fairness measures are not always compatible with one another, and thus, cannot be satisfied simultaneously \cite{chouldechova2017fair,friedler2016possibility}. Furthermore, Kearns et al.~\cite{kearns2017preventing} described the problem of fairness gerrymandering, in which a classifier may appear to be fair to each protected attribute separately, but not 
when defined over multiple protected attributes.  
Recent studies have attempted to address these limitations. For example, Kearns et al.~\cite{kearns2017preventing} and 
Hebert-Johnson et al.~\cite{kearns2017preventing} proposed methods to ensure the fairness definitions hold over an exponential or infinite class of subgroups. 

In addition to quantifying the notion of fairness, there has been considerable interest in developing fairness-aware machine learning algorithms. Current approaches can be divided into three categories. The first category includes method based on data preprocessing \cite{zemel2013learning,louizos2015variational,madras2018learning}. The rationale behind such methods is that the training data is the primary cause of bias in machine learning. For example, Zemel et al.~\cite{zemel2013learning} introduced an optimization algorithm to map data points into a new feature space to obfuscate information about group membership.  Louizos et al.~\cite{louizos2015variational} employed a variational autoencoder model to learn a representation that is invariant to the protected attributes while preserving as much information as possible. Madras et al.~\cite{madras2018learning} developed various formulations of adversarial representation learning for handling different types of group fairness measures while Masrour et al.~\cite{masrour2020bursting}  introduced a dyadic-level fairness criterion based on a modified network modularity measure and showed how it can be utilized for network link prediction tasks to overcome the filter bubble problem.

The issue of perceived fairness has been previously studied in social science and criminal justice disciplines
\cite{peiro2014,shulga2018}, in which the perception of fairness is associated with standards and social norms. In addition, there are various empirical studies focusing on understanding human's perceptions of fairness due to algorithmically-generated
decisions. For example, Woodruff et al. \cite{woodruff2018qualitative} conducted a qualitative study by interviewing individuals from several marginalized groups (based on race or class) in the United States and concluded that learning about algorithmic fairness evoked negative feelings that are connected to current national discussions about racial injustice and economic inequality. Moreover the participants of the study also revealed that algorithmic fairness could substantially affect their trust in a product or a company. In another study, Lee and Baykal \cite{lee2017algorithmic} investigated users' perception on algorithmic fairness. They concluded that decisions were viewed as being unfair when the algorithm’s assumptions of users did not account for multiple concepts of fairness, including cognitive and social behaviors in groups.

\section{Quantifying Fairness Perception}
Let $G=<V,E,X>$ be an attributed network, where $V$ is set of nodes (vertices), $E \subseteq V \times V$  is the set of links (edges), and $X \in \mathbb{R}^{|V| \times d}$ is the feature matrix associated with the nodes in the network. We assume $X$ can be further decomposed into $X = (X^{(p)}, X^{(u)})$, where $X^{(p)}$ are the protected attributes  and $X^{(u)}$ are the unprotected ones. The set of links can also be represented by an adjacency matrix, $A$, where $A_{ij}=1$ if a link exists between node $i$ and $j$. To simplify the discussion, we consider undirected networks in this study, though the ideas presented in this paper can be easily extended to directed networks. Furthermore, we denote $A^k = A^{k-1} \times A = \prod_{i=1}^k A$ as the self-product of the adjacency matrix. Thus, $A^{k}_{ij} > 0$ if there exists a path of length $k$ between nodes $i$ and $j$, and 0 otherwise.

We also assume that each node $v$ is associated with a target outcome, $y_v$. For brevity, we restrict the discussion here to binary outcomes $y_v \in \{0,1\}$ for the remainder of this paper. As an example, in the context of peer review network, each node corresponds to a submitted paper and links between papers are established if the two papers share the same authors or have authors who had previously collaborated with each other. The outcome $y_v$ of a given paper $v$ may indicate whether the paper is acceptable or unacceptable based on the reviews about its technical merits provided by the conference reviewers and meta-reviewers (if available).

We assume there exists a decision function $h: V \rightarrow \{0,1\}$ associated with each node in the network. Let $\mathcal{H}$ be the hypothesis space of all decision functions. Our goal is to learn a decision function $h \in \mathcal{H}$ that is consistent with the set of outcomes $Y$ while satisfying some fairness criterion. From the perspective of peer review network, the decision function $h$ may refer to the final decision whether to accept or reject the paper. Furthermore, we designate each node $v$ as follows depending on the consistency between the decision function $h(v)$ and its target outcome, $y_v$:
\begin{itemize}
    \item \textbf{True positive}: $y_v = h(v) = 1$
    \item \textbf{False positive}: $y_v = 0$, $h(v) = 1$
    \item \textbf{True negative}: $y_v = h(v) = 0$
    \item \textbf{False negative}: $y_v = 1$, $h(v) = 0$
\end{itemize}
In addition, the true positive rate and false positive rate of a decision function, $h$, can be computed as follows:
\begin{itemize}
    \item \textbf{True positive rate}, TPR = $\frac{\sum_v y_v h(v)}{\sum_v y_v}$
    \item \textbf{False positive rate}, FPR = $\frac{\sum_v (1-y_v) h(v)}{\sum_v y_v}$
\end{itemize}

Our goal is to determine how the final decisions are perceived by the individual nodes in the network. In particular, do they feel that the decisions are biased toward nodes that belong to certain groups (e.g., does the paper acceptance decision tend to favor those written by well-established authors or those from well-known institutions)? To answer this question, we assume each node $v$ is associated with a fairness perception function, $f(v,h)$, given a decision function $h$. The function provides a local, albeit myopic, view on the individual fairness of the nodes in a given network. 

\subsection{Axioms for Fairness Perception}

Before introducing our proposed fairness perception function, $f(v,h)$, we first outline the desirable properties of the function using the following set of axioms. We assume each node $v \in V$ is associated with the following tuple, $(v.X_p, v.X_u, y_v, N(v))$, where $v.X_p$ denotes the value of its protected attribute, $v.X_u$ denotes the value of its other (unprotected) attributes, $y_v$ denotes its target outcome, and $N(v)$ denotes its $\delta$-neighborhood, which is defined as follows:
\begin{equation}
   N(v) = \{u \ | \ \exists k \le \delta: A_{uv}^k > 0\}. 
   \label{eqn:neighbordef}
\end{equation}
For the remainder of our discussion, we assume $\delta = 1$, unless stated otherwise. Let $G_r = (V_r, E_r, X_r)$ be an ego-network for node $r$, where $V_r = N(r)$ is the 1-neighborhood of $r$, $E_r = \{(i,j) \ | \ i,j \in N(r) \ \textrm{and} \ (i,j) \in E\}$ and $X_r$ is the feature matrix associated with the protected and unprotected attributes for node $r$. We now present a set of axioms on the fairness perception function.
 
 \begin{enumerate}
     \item \textbf{Locality axiom}: If $h(v) = h'(v)$ and $\forall u \in N(v): h(u) = h'(u)$, where $h, h' \in \mathcal{H}$, then $f(v,h) = f(v,h')$.
     

    \item \textbf{Monotonicity axiom}: If $h(v) < h'(v)$ and $\forall u \in N(v): h(u) = h'(u)$, where $h, h' \in \mathcal{H}$, then $f(v,h) \le f(v,h')$.  
    \item  \textbf{Neighborhood expectation axiom}: If $h(v) = h'(v)$ and $\forall u \in N(v): h(u) \le h'(u)$, where $h, h' \in \mathcal{H}$, then $f(v,h) \ge \ f(v,h')$. 

   \item \textbf{Homogeneity axiom}: Let $G_u$ and $G_v$ be the induced sub-graph based on node sets $V_u = N(u\cup \{u\}$ and $V_v = N(v)\cup \{v\}$, respectively. If $G_u$ and $G_v$ are isomorphic with respect to the decision function $h$, then $f(u,h) = f(v,h)$. 
    
\end{enumerate}
For the last axiom, we say that a pair of networks, $G_r = V_r, E_r, X_r)$ and $G_s = (V_s, E_s, X_s)$, are isomorphic with respect to the decision function $h$ if there exists a bijection function $m: V_r\to V_s$ such that: 
\begin{itemize}
    \item $\forall u \in V_r: h(u) = h(m(u)), y_u = y_{m(u)}$ and $X_u = X_{m(u)} $. 
    \item $\forall (u_1,u_2) \in E_r: (m(u_1),m(u_2)) \in E_s$
\end{itemize}

The locality axiom states that the perception of fairness for an individual depends on the decision outcomes for other individuals in its neighborhood. As long as the outcomes for the node and its neighborhood remains unchanged, the fairness perception function should remain the same. 
The monotonicity axiom suggests that the perception of fairness for an individual never decreases if the decision changes in favor of the individual (assuming the decisions for its  neighbors remain unchanged). For example, if a previous decision on the paper was overturned (say from reject to accept), then one should expect the fairness perception to improve (or at least stays the same). In contrast, the neighborhood expectation axiom states that if the number of neighbors with favorable decisions increases, then fairness perception decreases monotonically. This is because, if more individuals in our reference group received favorable decisions, we expect the decision outcome to be favorable for us as well. The increased expectation makes it less likely for us to perceive the decision as fair if our paper is rejected. The fourth axiom is to ensure consistency of the fairness perception function when applied to different nodes in the network, The homogeneity axiom simply says that if two disparate nodes, including their respective neighborhoods, in the network receive similar decision outcomes, their perception of fairness should be the same. 

\subsection{Proposed Network-Centric Fairness Perception}
\begin{definition}[Network-Centric Fairness Perception]
Given a network $G = <V,E, X> $ and a decision function $h$, the network-centric fairness perception function is defined as follows:
\begin{equation}
    f(v, h) = \begin{cases} 1 & \text{if} \ \mathbb{E}[h(v)] \le h(v)\\ 0 & \text{otherwise} \\ \end{cases}
\label{eqn:netpercep}
\end{equation}
where $\mathbb{E}[h(v)]$ is the expected value of $h(v)$, which must satisfy the following properties:
\begin{enumerate}
    \item If $\forall u \in N(v): h(u) = h'(u)$, then $\mathbb{E}[h(v)] = \mathbb{E}[h'(v)]$.
    \item If $\forall u \in N(v): h(u) \le h'(u)$, then $\mathbb{E}[h(v)] \le \mathbb{E}[h'(v)]$.
    \item  Let $G_u$ and $G_v$ be the the induced sub-graphs based on the node sets $V_u = N(u) \cup \{u\}$ and $V_v = N(v) \cup \{v\}$, respectively. If $G_u$ and $G_v$ are  isomorphic with respect to the decision function $h$, then $\mathbb{E}[h(v)] = \mathbb{E}[h(u)]$.
\end{enumerate}
\end{definition}
Briefly speeaking, the function can be viewed as a local measure of individual fairness for any given node $v$ in a network. If the decision $h(v)$ is more favorable than expected, then $v$ will perceive the decision as fair. Furthermore, the expected value of the decision outcome, $\mathbb{E}[h(v)]$, depends on the neighborhood of the node $v$. 
\vspace{0.1cm}

\begin{theorem} 
The network-centric fairness perception function given in Eqn. \eqref{eqn:netpercep} satisfies the locality, monotonicity, neighborhood expectation, and homogeneity axioms. 
\end{theorem}

\emph{Proof}:
The locality and monotonicity properties can be easily proven using the first property. Since $\mathbb{E}[h(v)]$ remains unchanged when $h(u) = h'(u)$ for all the nodes $u$ in the neighborhood $N(v)$, Eqn. \eqref{eqn:netpercep} suggests that $f(v,h)$ depends only on $h(v)$. If $h(v) = h'(v)$, then $f(v,h)=f'(v,h)$, thereby proving that the locality axiom holds. Similarly, if $h(v) < h'(v)$, then $f(v,h) \le f(v,h')$, which satisfies the monotonicity axiom. For the neighborhood expectation axiom, the second property above states that the expected value monotonically decreases when $h(u) \le h'(u)$ for all nodes $u$ in the neighborhood $N(v)$. Since $\mathbb{E}[h'(v)]$ is larger, then nodes that initially satisfy the inequality $\mathbb{E}[h(v)] \le h(v)$ may no longer do so since $h'(v) = h(v)$. Thus, $f(v,h) \ge f(v,h')$. Finally, we use the the third property to prove the homogeneity axiom. Let $G_u$ and $G_v$ be the induced sub-graphs based on node sets $V_u = N(u)\cup \{u\}$ and $V_v = N(v)\cup \{v\}$, respectively. Since $G_u$ and $G_v$ are isomorphic with respect to the decision function $h$ 
and $\mathbb{E}[h(u)] = \mathbb{E}[h(v)]$ holds due to the third property, therefore $f(u,h) = f(v,h)$.
$\qed$

In principle, the network-centric fairness perception function can accommodate any definition of the expected value for the decision function $h(v)$, as long as it satisfies the 3 properties above. Here, we consider the following \textbf{neighborhood peer expectation} approach to compute $\mathbb{E}[h(v)]$:
$$\mathbb{E}[h(v)] = \frac{y_v}{k_1} \Big[\sum_{u \in N(v)} y_uh(u)\Big] 
+ \frac{1-y_v}{k_0} \Big[\sum_{u \in N(v)} (1-y_u)h(u)\Big],$$
where $k_0 = \sum_{u \in N(v)} (1-y_u)$, $k_1 = \sum_{u \in N(v)} y_u$, and $y_u \in \{0,1\}$.
Note that if the target outcome $y_v = 1$, then $\mathbf{E}[h(v)]$ depends only on the first term (i.e., other nodes $u$ in its neighborhood with $y_u = 1$). On the other hand, if $y_v = 0$, then  $\mathbf{E}[h(v)]$ depends only on the second term (i.e., other nodes $u$ in its neighborhood with $y_u = 0$). In the extreme case when the decision function assigns every node $v$ to $h(v)=0$, the expected value, $\mathbb{E}[h(v)]=0$ for all the nodes. Under this scenario, all the nodes perceive the decision to be fair since none of them are favored by the decision function. 

Intuitively, the neighborhood peer expectation considers the average decision of all its neighbors with the same target outcome. For example, if $y_u$ denotes whether paper $u$ is acceptable (based on its overall reviews) and $h(u)$ is its decision for acceptance, then the expected value of $h(u)$ depends on the average decision for other papers in its neighborhood (e.g., papers co-authored by one of the authors or their collaborators) with the same degree of  acceptability. Since the expectation is a monotonically increasing function of $h(u)$ for its neighbors, it can be trivially shown that the neighborhood peer expectation satisfies the first two properties given in Definition 1. The third property also holds since the bijection function guarantee the $y$ and $h$ values for the neighborhoods of $u$ and $v$ are exactly the same and as a results their expectation is going to be the same.    

\section{Fairness Visibility}
The fairness perception function described in the previous section is a local measure for individual fairness by considering the decision outcomes of a node as well its neighbors. In this section, we introduce the notion of fairness visibility, which extends the fairness perception function to a group fairness measure.

\begin{definition}[Fairness Visibility]
Let $V_c = \{u \ | \ u \in V, \ u.X_p = c\}$, i.e., the set of nodes belonging to the protected group $c$. The fairness visibility of $h$ for group $c$ is defined as follows:
\begin{equation}
    FV(V_c) = \frac{\sum_{v \in V_c}f(v,h)}{|V_c|}
  \end{equation} 
\end{definition}
Note that the fairness visibility for a given group $c$ can be viewed as the average fairness perception of all the nodes that belong to the protected group $c$. For example, the group $c$ may refer to all the papers written by well-established authors in the peer review network. To determine whether the decision function $h$ is fair, we may compare the fairness visibility for different groups of nodes using the definition below.

\begin{definition}[Fairness Visibility Parity] The decision function $h$ satisfies fairness visibility parity for $V_c$ and $V_{c'}$ if 
\begin{equation}
    FV(V_c) = FV(V_{c'})
\end{equation}
\end{definition}
For example, in a peer review network, we may categorize the papers into two groups, those written by famous authors or those written by less established researchers. If the average fairness perception for both groups of papers are the same, then their fairness visibility parity holds. The larger the disparity, the more biased are the decisions (based on the perception of the nodes in the network). 

A standard approach for measuring group fairness is to compute demographic parity, which is defined as follows:
\begin{definition}[Demographic Parity] The decision function $h$ satisfies demographic parity for $V_c$ and $V_c'$ if
\begin{equation}
    P(h(v) = 1 \ | \ v \in V_c) = P(h(v) = 1 \ | \ v \in V_c') 
\label{eqn:demo}
\end{equation}
\end{definition}
Unlike fairness visibility parity, demographic parity is computed for non-relational data since it ignores the neighborhood structure of a node. In the context of peer review network, each probability term in Eqn. \eqref{eqn:demo} corresponds to the acceptance rate of papers that belong to the group $c$ or $c'$. For brevity, we termed $P(h(v)=1 | v \in V_c)$ as the acceptance probability for the group $V_c$. In the theorem below, we illustrate the relationship between fairness visibility $FC(V_c)$ and acceptance probability for $V_c$. Specifically, we show that the fairness visibility measure reduces to acceptance probability under certain mild conditions.
\vspace{0.1cm}

\begin{theorem} 
\label{th:converge}
Assuming the network graph is connected and the decision function $h$ has non-zero true positive and false positive rates, the fairness visibility of group $V_c$, based on the neighborhood peer expectation, converges to the acceptance probability for $V_c$ as the $\delta$-neighborhood size increases. 
\end{theorem}
\vspace{0.1cm}

\emph{Proof}:
Given a node $v$, note that $N(v) \rightarrow V$ as the $\delta$-neighborhood expands since the network graph is assumed to be connected. Furthermore, if the true positive and false positive rates for $h$ are non-zeros, then $\mathbb{E}[h(v)] > 0, \forall v \in V$. It follows that $f(v,h) =1$ if $h(v) =1$ and $f(v,h) = 0$ if  $h(v) =0$. Thus $FV(V_c)$ converges to $P(h(v) = 1|v \in V_c)$.
$\qed$
\vspace{0.1cm}

\begin{corollary} Given a connected network $G$, the decision function $h$ satisfies demographic parity if and only if there exists a positive integer $k>0$ such that for all $\delta \ge k$, fairness visibility parity is satisfied for $h$ on $G$ with the given $\delta$-neighborhood.  
\end{corollary}

\section{Application to Peer Review Networks}
\label{sec-Exp}
This section presents a case study on the application of the proposed network-centric  fairness perception and fairness visibility measures on a peer review network dataset. Specifically, these measures allow us to assess how fair are the paper acceptance decisions when viewed from a local neighborhood perspective. 

\begin{figure}[t!]
    \centering
    \includegraphics[width=\linewidth]{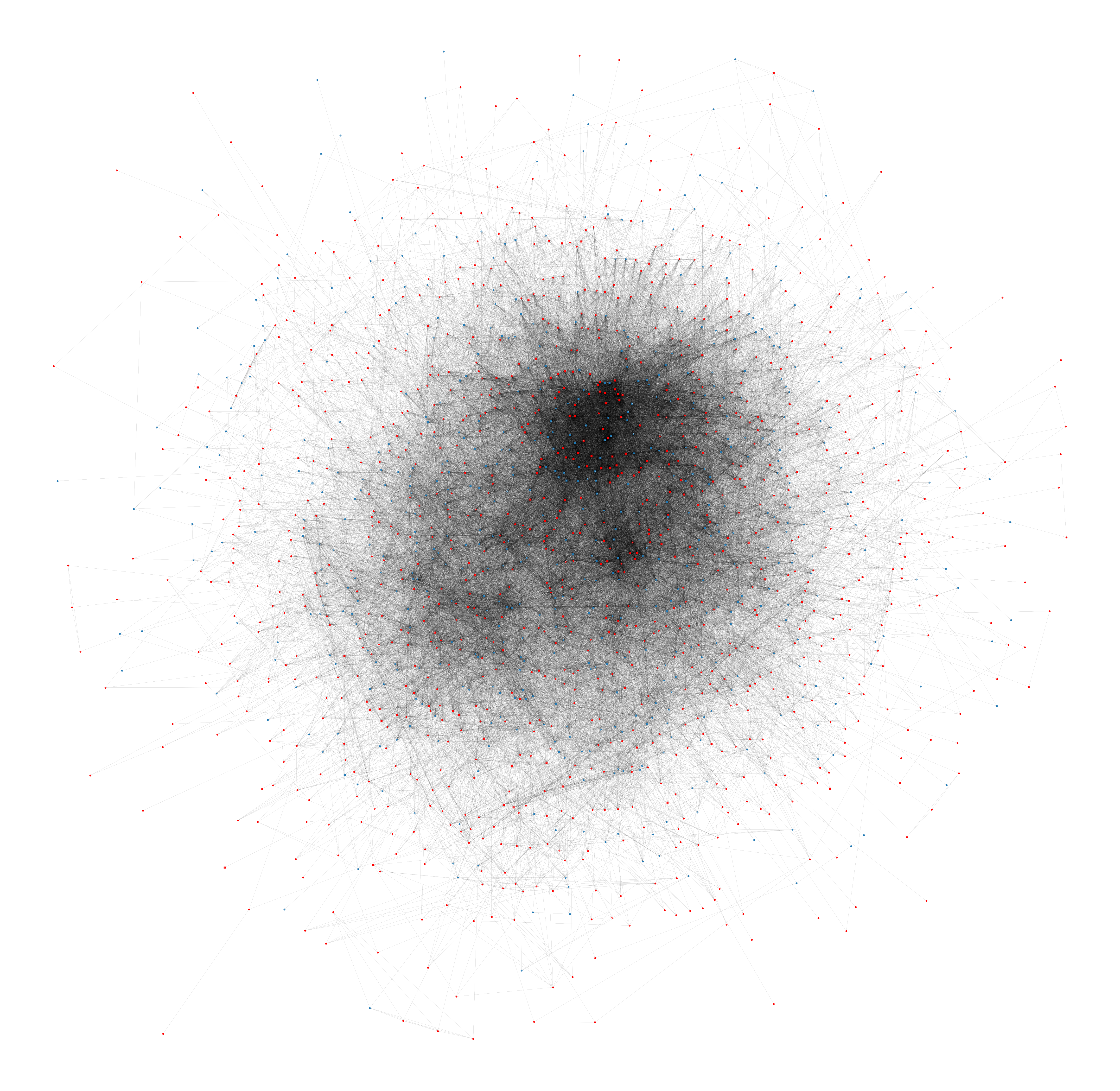}
    \caption{Peer review network of papers submitted to the ICLR 2020 conference. The nodes in the network correspond to the submitted papers and the links are established based on co-author relationships.}
    \label{fig:peer_network}
\end{figure}

\subsection{Data}
We constructed a network from the
peer review dataset collected for the ICLR 2020 conference from the \url{OpenReview.net} website. The website provides detailed information on the peer review evaluation process for various workshops and conferences (mostly in computer science). Specifically, for each submitted paper, we gathered information about its title, abstract, list of authors and their affiliations. In addition, the anonymized reviews and acceptance decision for each reviewed paper are also available. For the ICLR 2020 conference, the number of submitted papers is 2594. However, 382 of the submissions were withdrawn. Our analysis is therefore restricted to only 2212 papers which had been reviewed. We use this information to create a network that contains 2212 nodes, one for each peer-reviewed paper. 
\begin{figure}
    \centering
    \includegraphics[width=1\linewidth]{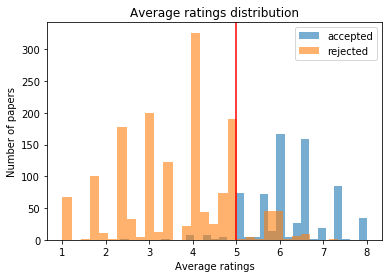}
    \caption{The average rating distribution for accepted and rejected papers submitted to the conference, the red line indicate the threshold used for classifying papers as acceptable ($y = 1$) or unacceptable ($y = 0$).}
    \label{fig:his_avg_rating}
\end{figure}

The total number of accepted papers, either as oral or poster presentation, is 687 while the number of rejected papers is 1525. Thus, the acceptance rate for the conference is around 31\%. We use the acceptance decision of each paper as the decision function $h$ to be evaluated in terms of its fairness perception. We also consider the acceptability of the paper in terms of its average review ratings as our target outcome $y$. Our assumption here is that the reviewers are rational-minded individuals, whose average ratings given to a paper reflect the technical merits and acceptability level of the paper. Figure \ref{fig:his_avg_rating} shows histograms of average review ratings for the accepted and rejected papers. Given that the number of accepted papers is 687, 
we choose an acceptability threshold of 6 since it gives a number of acceptable papers that has the closest match to the number of accepted papers. With this threshold, all papers whose average ratings are larger than 5 are considered acceptable, i.e., $y=1$. 
Table \ref{tab:genera_conf}(a) shows a confusion matrix comparing acceptability of the paper ($y$) and its acceptance decision ($h$). 


\begin{table}[]
\centering
\caption{Summary distribution of acceptable and accepted papers for the ICLR 2020 conference.}
    \label{tab:genera_conf}
\begin{tabular}{c|l|c|c|}
\multicolumn{2}{c}{}&\multicolumn{2}{c}{Acceptability}\\
\cline{3-4}
\multicolumn{2}{c|}{}& $y=1$ & $y=0$ \\
\cline{2-4}
Acceptance & $h=1$ & 589 & 98 \\
\cline{2-4}
Decision & $h=0$ & 117 & 1408 \\
\cline{2-4} 
\multicolumn{4}{c}{ \ }\\ 
\multicolumn{4}{c}{(a) All papers}
\end{tabular}
\begin{tabular}{c|l|c|c|}
\multicolumn{2}{c}{}&\multicolumn{2}{c}{Acceptability}\\
\cline{3-4}
\multicolumn{2}{c|}{}& $y=1$ & $y=0$ \\
\cline{2-4}
Acceptance & $h=1$ & 94 & 13 \\
\cline{2-4}
Decision & $h=0$ & 12 & 153 \\
\cline{2-4}
\multicolumn{4}{c}{ \ }\\ 
\multicolumn{4}{c}{(b) Famous authors}\\
\multicolumn{4}{c}{ \ }\\ 
\end{tabular}
\begin{tabular}{c|l|c|c|}
\multicolumn{2}{c}{}&\multicolumn{2}{c}{Acceptability}\\
\cline{3-4}
\multicolumn{2}{c|}{}& $y=1$ & $y=0$ \\
\cline{2-4}
Acceptance & $h=1$ & 495 & 85  \\
\cline{2-4}
Decision & $h=0$ & 105 & 1255 \\
\cline{2-4}
\multicolumn{4}{c}{ \ }\\ 
\multicolumn{4}{c}{(c) Non-famous authors}\\
\multicolumn{4}{c}{ \ }\\ 
\end{tabular}
\begin{tabular}{c|l|c|c|}
\multicolumn{2}{c}{}&\multicolumn{2}{c}{Acceptability}\\
\cline{3-4}
\multicolumn{2}{c|}{}& $y=1$ & $y=0$ \\
\cline{2-4}
Acceptance & $h=1$ & 190 & 34 \\
\cline{2-4}
Decision & $h=0$ & 21 & 328 \\
\cline{2-4}
\multicolumn{4}{c}{ \ }\\ 
\multicolumn{4}{c}{(d) Top institutions}\\
\multicolumn{4}{c}{ \ }\\ 
\end{tabular}
\begin{tabular}{c|l|c|c|}
\multicolumn{2}{c}{}&\multicolumn{2}{c}{Acceptability}\\
\cline{3-4}
\multicolumn{2}{c|}{}& $y=1$ & $y=0$ \\
\cline{2-4}
Acceptance & $h=1$ & 399 & 64  \\
\cline{2-4}
Decision & $h=0$ & 96 & 1080 \\
\cline{2-4}
\multicolumn{4}{c}{ \ }\\ 
\multicolumn{4}{c}{(e) Non-top institutions}
\end{tabular}
\end{table}
 
The total number of authors who had submitted papers to the conference was 6953. We were able to extract authorship information for each paper, such as names and email addresses of the co-authors, affiliation, gender, and scholarid by prepossessing the the users profile page on the OpenReview website. Based on this information we classified the submitted papers into groups based on the following ``protected" attributes:
 \begin{itemize}
    \item \textbf{Famous author papers}: If a paper includes one or more famous authors, its protected attribute value is $X_p=0$, otherwise $X_p=1$. We used a list provided by Google Scholar to identify the famous authors based on their h-index rankings\footnote{\url{https://scholar.google.com/citations?view_op=search_authors&hl=en&mauthors=label:machine_learning}}. 
    We consider the top 500 authors in the list as famous authors. With this designation, 272 of the submitted papers were classified as co-authored by famous authors.
    \item \textbf{Top institution papers}:  If a paper has an author from a top university, then it its protected attribute value is $X_p = 0$, otherwise $X_p=1$. To do this, we consider only authors from the top 10 universities according to the csrankings.org website\footnote{\url{http://csrankings.org/#/index?all}}.  With this designation, 573 of the submitted papers have at least one author from the top institution.
\end{itemize}

A breakdown on the number of acceptable and accepted or rejected papers by each protected group is shown in Tables \ref{tab:genera_conf}(b)-(e). The results shown in these tables are consistent with previous research, which had suggested that conference paper acceptance decisions are generally biased in favor of famous authors or papers written by authors from top universities~\cite{stelmakh2018peerreview4all,tomkins2017reviewer}.  
In particular, the results suggest that the chance of an acceptable paper by a famous author or an author from top institution being accepted is significantly higher. For example, comparing Tables \ref{tab:genera_conf}(b) and (c), we observe that the true positive rate for papers by famous authors (88.7\%) is significantly higher than that for those by non-famous authors (82.5\%). Similarly, comparing Tables \ref{tab:genera_conf}(d) and (e), papers by authors from top institutions also have a higher true positive rate (90.1\%) compared to those written by authors from lower ranked institutions (80.6\%).  

Papers by famous or top institution authors also have a higher chance of getting their unacceptable papers accepted compared to those written by non-famous authors or authors from lower ranked institutions. This is reflected by the higher false positive rates in the tables. For example, papers by famous authors have a higher false positive rate (7.8\%) than those by non-famous authors (6.3\%) whereas papers by authors from top institutions have a significantly higher false positive rate (9.4\%) compared to those from lower ranked institutions (5.6\%).
\begin{figure}
    \centering
    \includegraphics[width=1\linewidth]{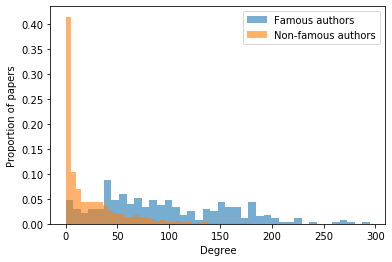}
    \caption{Degree distribution for papers in the peer review network based on the famous author protected attribute.}
    \label{fig:famous_degree_dist_top}
\end{figure}

\begin{figure}
    \centering
    \includegraphics[width=1\linewidth]{figs/2020_hist_degree_top.png}
    \caption{Degree distribution for papers in the peer-review network based on the top institution protected attribute.}
    \label{fig:famous_degree_dist}
\end{figure}

We use the co-authorship information extracted from the authors' profile pages on OpenReview.net to construct the links between the nodes in the network. We consider two papers are linked if they share a common co-author or if the authors have collaborated in the past. Based on this approach, the resulting network contains 36,097 links, with an average degree around 16. Figure \ref{fig:famous_degree_dist} shows the degree distribution of the network based on the famous author protected attribute. The results suggest that papers by famous authors tend to have higher degree (on average) and a flatter degree distribution compared to those written by non-famous authors. Similarly, Figure \ref{fig:famous_degree_dist_top} shows the degree distribution of the network based on the top institution protected attribute. Once again, the results show that papers written by authors whose protected attribute value is $X_p = 0$ (i.e., authors from Top institutions) tend to have higher degree and a flatter distribution. 



\subsection{Fairness Perception}
\begin{figure}
    \centering
    \includegraphics[width=1\linewidth]{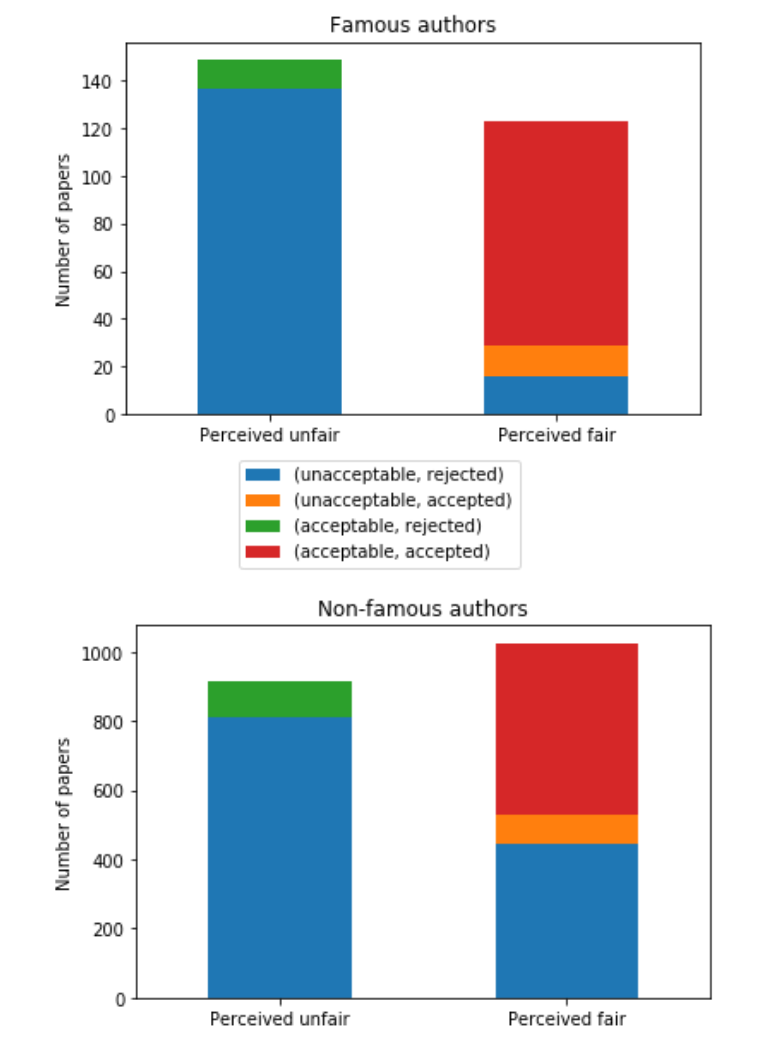}
    \caption{Assessment of network-centric fairness perception based on the famous author protected attribute.}
    \label{fig:fairness_perc}
\end{figure}

\begin{figure}
    \centering
    \includegraphics[width=1\linewidth]{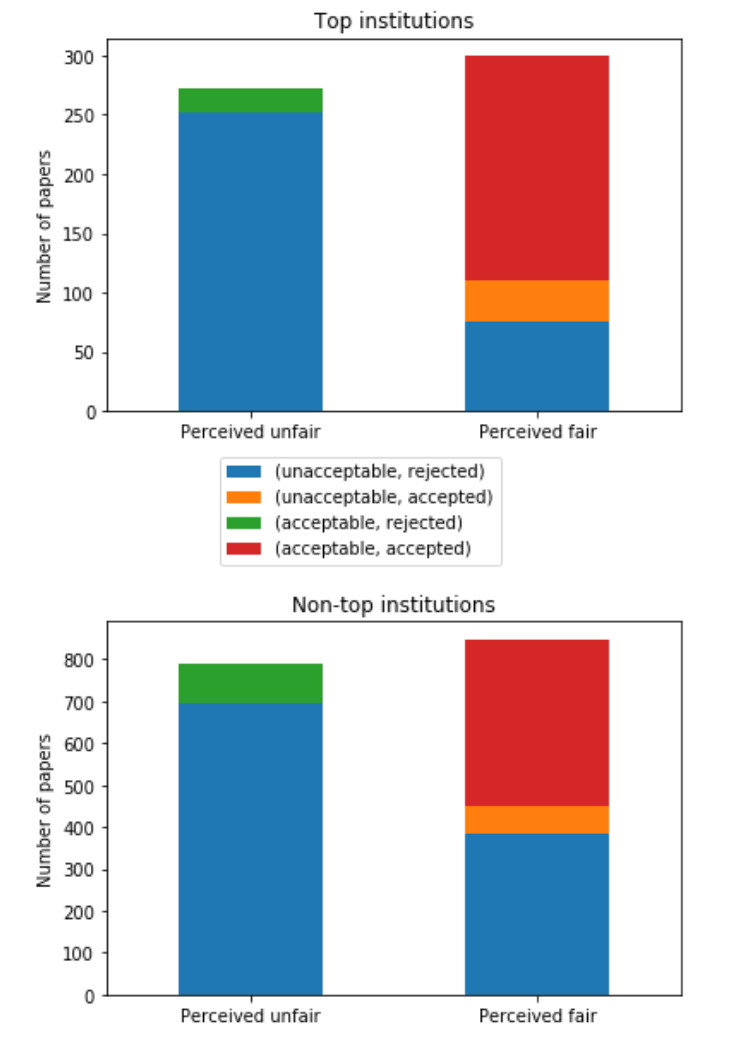}
    \caption{Assessment of network-centric fairness perception based on the top institution protected attribute.}
    \label{fig:fairness_perc}
\end{figure}

\begin{figure}
    \centering
    \includegraphics[width=1\linewidth]{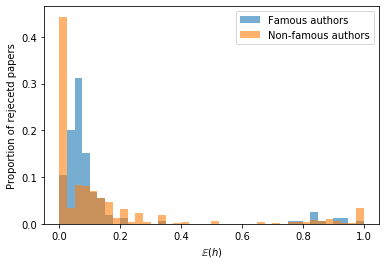}
    \caption{Comparison of $\mathbb{E}[h(v)]$ for rejected papers by famous and non-famous authors.}
    \label{fig:exp_famous_rej}
\end{figure}

We applied the network-centric fairness perception function to the network and evaluated the proportion of papers who perceived the paper acceptance decision to be fair or unfair. The results are shown in  Figure~\ref{fig:fairness_perc} for the famous author protected attribute. Despite the fact that papers by famous authors are generally favored (i.e., have higher true positive and false positive rates), the bar chart shown in Figure~\ref{fig:fairness_perc}(top) suggests that the majority of them perceived the decision to be unfair. According to the network-centric fairness perception function, their main source of dissatisfaction is due to unacceptable papers that were rejected (i.e., the blue bar), which they believe should be accepted.  For papers by non-famous authors, the bar chart shown in Figure~\ref{fig:fairness_perc}(bottom) tells an opposite story. Here, the majority of the papers perceived the paper acceptance decisions to be fair. Although there is still a significantly large proportion of unacceptable papers that were rejected categorized as perceived unfair (i.e., the blue bar), the papers by non-famous authors are also more amenable to accepting the decision to reject their unacceptable papers (see the blue bar for perceived fair). 


The preceding results show a potential pitfall of using the fairness perception function (with neighborhood size $\delta = 1$). Although the analysis of the confusion matrices given in Table \ref{tab:genera_conf} suggests that the decision is biased in favor of papers with $X_p = 0$ (famous authors or top institutions), those with $X_p = 1$ still perceived the decisions to be fair! 
We will explain the reason for the observed discrepancy in the remainder of this section and provide an approach to alleviate this problem in the next section. 

Since our network-centric fairness perception function depends on  computation of $\mathbb{E}[h(v)]$, we examine the distribution of the expected values for the rejected papers by famous and non-famous authors. The results are shown in  Figure \ref{fig:exp_famous_rej}. 
Our first observation is that more than $40\%$ of the rejected papers by non-famous authors have an expected value close to 0 compared to around 10\% of the rejected papers by famous authors, 
Based on the definition given in Eqn. \eqref{eqn:netpercep}, the larger the proportion of papers with $\mathbb{E}[h(v)]$ close to zero, the more likely they perceived the decision to be fair. 
One possible explanation for the famous authors to have fewer proportion of papers with $\mathbb{E}[h(v)]$ close to zero can be attributed to the degree distribution of their nodes (see Figure \ref{fig:famous_degree_dist_top}). Since papers by famous authors generally have a higher degree, this increases the number of nodes in their neighborhood, which in turn, results in a higher expected value according to the formula used to compute the neighborhood peer expectation. In contrast, many of the papers by non-famous authors have low degree nodes, which in turn, produces more nodes with low $\mathbb{E}[h(v)]$.   


\subsection{Fairness Visibility}
We introduced fairness visibility as an extension of our fairness perception function to group fairness measure. Theorem 2 and Corollary 2.1 also show its relationship to demographic parity under certain mild assumptions. The results in the previous section also suggested a potential pitfall of assessing fairness from a local neighborhood perspective since the perceived fairness for papers with $X_p=1$ is higher than those with $X_p=0$ even though the decisions are actually biased against them. 
In this section we will experimentally evaluate the theoretical results for fairness visibility, which provides a possible solution to alleviate the potential pitfall of using our fairness perception function. 

\begin{figure} 
    \centering
  \subfloat[Famous author]{%
       \includegraphics[width=0.8\linewidth]{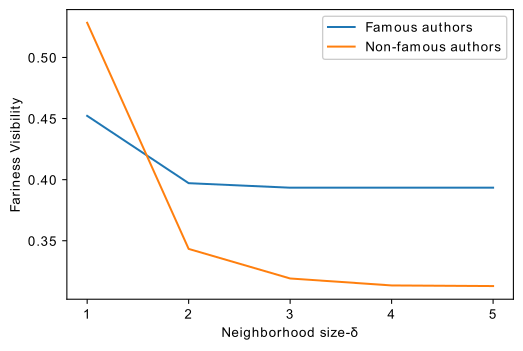}}
    \hfill
  \subfloat[Top institution]{%
        \includegraphics[width=0.8\linewidth]{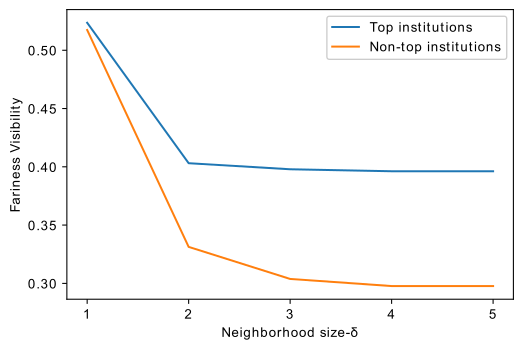}}
  \caption{Effect of neighborhood size on fairness visibility.}
  \label{fig:vis_expand} 
  \vspace{-0.3cm}
\end{figure}

For Figure \ref{fig:vis_expand}-(a), observe that the fairness visibility of papers by famous authors are initially lower than that for papers by non-famous authors when $\delta=1$. This means that, on average, the papers by famous authors have lower perceived fairness. As $\delta$ increases, fairness visibility decreases for both groups of papers. However, the rate of decrease is higher for papers by non-famous authors. According to Theorem 2, under mild assumption, fairness visibility will converge to the acceptance probability of each subgroup of the protected attribute when $\delta$ increases. Since the acceptance probability for $X_p=0$ (famous authors) is higher than that for $X_p=1$ (non-famous authors), the fairness visibility for famous authors will be higher for larger values of the neighborhood size, $\delta$. This provides a strategy to counter against the potential pitfall of using fairness perception by expanding the neighborhood size $\delta$. Furthermore, it is worth noting that the peer review network is not a connected graph. As a result, the fairness visibility does not converge exactly to the acceptance probability for each group, which is 0.2989 (for non-famous authors) and 0.3933 (for famous authors),  when $\delta$ is sufficiently large.


 A similar observation can be made when analyzing the effect of increasing neighborhood size on fairness visibility using top institution as protected attribute. As shown in Figure \ref{fig:vis_expand}-(b), increasing $\delta$ leads to lower fairness visibility. However, with sufficiently large $\delta$, the fairness visibility for papers by authors from top institutions is higher than that for papers by authors from lower ranked institutions. By setting $\delta=2$, the fairness visibility provides a good assessment on the true bias of the paper acceptance decisions. 



\section{Conclusion}
This paper presents a novel approach for algorithmic fairness in network data. Motivated by the equity theory in social science, we introduced the concept of fairness perception as a local formulation of fairness and quantified this notion through an axiomatic approach to analyze its properties. We also showed how our proposed network-centric fairness perception function can be extended to a group fairness measure known as fairness visibility. We provided theoretical analysis to demonstrate its relationship to demographic parity. Using a peer-review network as case study, we also examined its utility in terms of assessing the perception of fairness in paper acceptance decisions. We also highlighted a potential pitfall of using fairness visibility measure as it can be exploited to mislead individuals into perceiving that the algorithmic decisions are fair. Finally, we show how to alleviate the problem by increasing the local neighborhood size. 

For future work, we plan to apply the fairness visibility measure for node classification and other network analysis tasks. We will also extend the approach to handle other types of networks (e.g., k-partite networks consisting of heterogeneous nodes).


\section*{Acknowledgment}

This work is supported in part by the U.S. National Science Foundation under grant \#IIS-1939368.

\bibliography{fairness}
\bibliographystyle{IEEEtran}
%
%

\end{document}